\documentclass[twocolumn,showpacs,aps,prd,nobibnotes,
  nofootinbib,floatfix,superscriptaddress,longbibliography]{revtex4-1}

\usepackage{natbib}
\usepackage{amsmath,amssymb,bm,xfrac} 
\usepackage{graphicx}
\usepackage[usenames]{xcolor}
\usepackage[normalem]{ulem}
\usepackage{xspace}
\usepackage{dsfont} 
\usepackage{soul} 
\usepackage{dsfont} 
\usepackage{multirow}  
\usepackage{float}  
\usepackage{cancel}
\usepackage[caption=false]{subfig}
\usepackage[colorlinks=true,allcolors=blue]{hyperref} 
\usepackage[caption=false]{subfig}

\newcommand{\orcid}[1]{\href{https://orcid.org/#1}{\includegraphics[height=\fontcharht\font`\B]{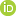}}}

\newcommand{\apjs}{Astrophys. J. Lett.}
\newcommand{\aap}{Astron. Astrophys.}

\newcommand{\apjl}{Astrophys. J. Lett.}

\newcommand{\jcap}{J.~Cosmology Astropart. Phys.}

\newcommand{\mnras}{Mon. Notices Royal Astron. Soc.}

\newcommand{\pasp}{Publ. Astron. Soc. Pac.}

\newcommand{\beq}{\begin{equation}}
\newcommand{\eeq}{\end{equation}}

\renewcommand{\eqref}[1]{Eq.\,(\ref{#1})\xspace}

\definecolor{darkred}{cmyk}{0,1,1,0.4}

\begin{document} 

    \title{Joint Analysis of Constraints on $f(R)$ Parametrization from Recent Cosmological Observations}

\author{Darshan Kumar~\orcid{0000-0001-6665-8284}}
\email{kumardarshan@hnas.ac.cn, darshanbeniwal11@gmail.com}
\affiliation{Institute for Gravitational Wave Astronomy, Henan Academy of Sciences, Zhengzhou, Henan 450046, China}
 
\author{Praveen Kumar Dhankar~\orcid{0000-0002-8201-6019}}     
\email{pkumar6743@gmail.com, praveen.dhankar@sitnagpur.siu.edu.in}
\affiliation{Symbiosis Institute of Technology, Nagpur Campus, Symbiosis International (Deemed University), Pune 440008, Maharastra, India}

\author{Saibal Ray~\orcid{0000-0002-5909-0544}}
\email{saibal.ray@gla.ac.in}
\affiliation{Centre of Cosmology, Astrophysics and Space Science (CCASS), GLA University, Mathura 281406, Uttar Pradesh, India}    
      
\author{Fengge Zhang~\orcid{0000-0001-7365-5057}}
\email{zhangfengge@hnas.ac.cn}
\affiliation{Institute for Gravitational Wave Astronomy, Henan Academy of Sciences, Zhengzhou, Henan 450046, China}
 
%


\begin{abstract}
In this study, we present constraints on the parameters of three well-known $f(R)$ gravity models, viz. (i) Hu-Sawicki, (ii) Starobinsky, and (iii) ArcTanh by using a joint analysis of recent cosmological observations. We perform analytical approximations for the Hubble parameter, $H(z)$, and cosmological distances in terms of the Hubble constant $(H_0)$, matter density $(\Omega_{m0})$, and a deviation parameter $b$ for each model. {Our analysis combines early and late-universe cosmological data from five cosmological observations:} (a) Hubble parameter measurements (Cosmic Chronometers), (b) Type Ia Supernovae (Union 3.0), (c) Baryon Acoustic Oscillations (DESI-2025), (d) Gamma-Ray Bursts (GRBs) and (e) Cosmic Microwave Background (CMB). We first optimize the models using each dataset independently, and subsequently, we perform a comprehensive joint analysis combining all four datasets. Our results show that the Hu-Sawicki and ArcTanh models do not deviate significantly from the $\Lambda$CDM model at 95\% confidence level for individual datasets and remain consistent at 99\% confidence level in the joint analysis. In contrast, the Starobinsky model shows a strong deviation and appears as a viable alternative to $\Lambda$CDM. We also constrain the transition redshift parameter ($z_t$), and check that the obtained value agrees with the values inferred from both early-time measurement (Planck) and late-time data from Type Ia Supernovae. These results support the potential support of $f(R)$ gravity to explain the late-time cosmic acceleration effectively. Finally, a statistical model comparison using $\chi^2_{\text{min}}$, AIC, and BIC indicates that all three $f(R)$ models are favored over $\Lambda$CDM, with the Starobinsky model receiving very strong support.\\
	
\end{abstract}

\maketitle

\section{Introduction}        
The late-time acceleration of the universe remains one of the most significant challenges in modern cosmology. The standard $\Lambda$CDM model, which attributes cosmic acceleration to a cosmological constant ($\Lambda$) and cold dark matter (CDM), provides an excellent fit to a wide range of cosmological observations. However, it suffers from theoretical issues such as the fine-tuning problem and the cosmic coincidence problem \citep{2000astro.ph..5265W,2003RvMP...75..559P}, as well as observationally from the mismatch between local and global measurements of the Hubble constant. The Hubble constant $H_0$ has been used for almost $90$ years to describe the current rate of expansion of the Universe \cite{Lemaitre:1927zz,Hubble:1929ig}. Although, while the precise measurement of the Cosmic Microwave Background Radiation (CMBR) combined with the standard ($\Lambda$CDM) cosmological model indicate that $H_0 = 67.66 \pm 0.42$ $\mathrm{km} \ \mathrm{s}^{-1}\mathrm{Mpc}^{-1}$ \citep{WMAP:2012nax,2020A&A...641A...6P}, a model-independent calibration through distance ladders using Type Ia supernovae (SNe) data and the SH0ES project produces $H_0 = 73.04 \pm 1.04$ $\mathrm{km} \ \mathrm{s}^{-1}\mathrm{Mpc}^{-1}$ \cite{Pan-STARRS1:2017jku,Riess:2021jrx}. This significant (nearly $5 \sigma$) conflict between the outcomes of two profoundly different measurements (early time vs. local) is termed as the {\it Hubble tension}. For comprehensive reviews of the Hubble tension phenomenology, interested readers can check out the following refs.~\citep{2021APh...13102605D,2021ApJ...919...16F,2020PhRvD.101d3533K,2022JHEAp..34...49A,2025arXiv250401669D}. In addition, the Standard Cosmological model shows inconsistencies with the observed amplitude of matter fluctuations ($S_8$) \cite{2020A&A...641A...6P,2023OJAp....6E..36D} and the measured sound horizon at the time of Baryon Acoustic Oscillations (BAO) \cite{2022PhRvD.105d3528G,2022LRR....25....6M}. 

These cosmological tensions have motivated various theoretical proposals to resolve the apparent discrepancies. Some approaches invoke exotic dark energy components like quintessence or $k$-essence fields \citep{2005PhRvL..95n1301C,2013CQGra..30u4003T,2016ARNPS..66...95J,2023ForPh..7100133M,2023Univ....9..510M,2024ChJPh..91..838P,2024IJMPD..3350005M},\footnote{Remarkably, which would need to dominate the universe's energy budget at $\sim$70\%} while others attempt to reconcile vacuum energy contributions from quantum field theory. However, such quantum vacuum calculations yield predictions that diverge from observations by up to 120 orders of magnitude \citep{1995Natur.377..600O,1999Sci...284.1481B}, representing perhaps the most severe theoretical discrepancy in modern physics. 

On the other hand, the most direct approaches modify the dark energy equation of state through parameterizations like the Chevallier-Polarski-Linder form $w(z) = w_0 + w_a z/(1+z)$, though introducing new degeneracies \citep{2001IJMPD..10..213C,2003PhRvL..90i1301L,2021MNRAS.501.5845Y}. More fundamentally, the Lemaître-Tolman-Bondi solutions \citep{1999MPLA...14.1539P,2000A&A...353...63C}, propose that large-scale inhomogeneities could mimic dark energy effects without exotic components, though these models require fine-tuned observer positions and face increasing observational constraints.   

Beyond dark energy parameterizations and inhomogeneous models, modified gravity theories provide a geometrically alternative approach by introducing higher-order corrections to the Einstein-Hilbert action. The simplest modification of general relativity replaces the Ricci scalar $R$ with a generic function $f(R)$, originally proposed for inflation but now widely studied for late-time acceleration \citep{2012PhR...513....1C,2010LRR....13....3D}. These theories generate cosmic acceleration through spacetime dynamics itself, avoiding explicit dark energy components while requiring careful tuning to satisfy both cosmological observations and local gravity tests \citep{2007PhRvD..75h3504A,2010RvMP...82..451S,2011PhR...505...59N,2007IJGMM..04..115N,2017PhR...692....1N,2003PhRvD..68l3512N}. Various functional forms of $f(R)$ have been proposed to achieve these goals while recovering the $\Lambda$CDM model in certain limits and other cosmological aspects \citep{Carroll2004, Mishra2021}. 

Current research focuses on identifying viable $f(R)$ functions that simultaneously: (i) reproduce $\Lambda$CDM's successes, (ii) satisfy solar-system tests through the chameleon mechanism, and (iii) predict distinguishable signatures in large-scale structure formation \citep{2001PhLB..506...13H,2007PhRvD..75d4004S}. The challenge persists in finding theoretically motivated forms that pass increasingly precise observational constraints while maintaining mathematical consistency. For example, early models like $f(R)=R+\alpha R^2$ successfully described inflation \citep{1980PhLB...91...99S}, other attempts such as $f(R)=1/R$ \cite{2004PhRvD..70d3528C} for late-time acceleration were quickly ruled out due to instabilities and solar-system constraints \citep{2003PhLB..573....1D,2003PhRvD..68l3512N,2003PhLB..575....1C}. Modern viable models must simultaneously satisfy cosmological observations and local gravity tests, with successful examples including the Hu-Sawicki \citep{2007PhRvD..76f4004H}, Starobinsky \citep{1980PhLB...91...99S}, Tsujikawa \citep{2008PhRvD..77b3507T}, and Exponential \citep{2009PhRvD..80l3528L} formulations. These models demonstrate $f(R)$ gravity's capacity to unify inflation and cosmic acceleration while remaining observationally viable. 

Observational constraints on $f(R)$ gravity have been actively investigated through multiple cosmological probes. Early analyses focused on the Hu--Sawicki model, using X-ray cluster gas mass fractions combined with CMB/BAO ratios and Union 2.1 supernovae \citep{2012A&A...548A..31S}, while the exponential model was first constrained with Union 2 supernovae and cosmic chronometer data \citep{2011PhLB..699..320C}.

{In order to investigate the modified gravity models at high redshift, a careful analysis of early universe observables is required. Cosmic Microwave Background (CMB) data from the Planck 2018 release provides particularly powerful constraints, through both temperature and polarization power spectra \cite{2020A&A...641A...6P}. These measurements allow precise tests of deviations from $\Lambda$CDM cosmology by probing the gravitational potential evolution across cosmic time. Recent analyses incorporating CMB distance priors, such as the shift parameter and the acoustic scale, provide additional leverage in breaking degeneracies between background expansion history and growth of structure \cite{2019JCAP...02..028C,2020JCAP...07..009Z}. This combination proves particularly effective for testing non-minimal couplings and $f(R)$ modifications. Subsequent studies have systematically expanded the scope of tested gravitational theories. The model space now includes both power-law ($R - \beta/R^n$) and logarithmic ($R + \alpha\ln R - \beta$) formulations \citep{2006A&A...454..707A,2007PhRvD..75f3509F,2008JCAP...09..008C}. These investigations increasingly combine diverse cosmological probes, ranging from baryon acoustic oscillation measurements to refined CMB shift parameters, offering comprehensive tests of modified gravity scenarios.

}

The field matured through comprehensive multi-model analyses incorporating growth rate measurements, local $H_0$ determinations, and Type Ia supernovae \citep{2013PhRvD..87l3529B,2017JCAP...01..005N,2018PhRvD..97b3525P,Odintsov:2018qug}. Recent work has further advanced the constraints by introducing three innovations: (i) gravitational lensing systematics \citep{2019PhRvD.100d4041D,2020PhRvD.101j3505D,2021NuPhB.96615377O}, (ii) quasar X-ray/UV flux ratios \citep{2022PhRvD.105j3526L}, (iii) HII galaxies \cite{10.1093/mnras/stad3705,2022MNRAS.514.5827S,2024arXiv240504886R}, (iv) weak lensing surveys \citep{2024arXiv241204807B}, (iv) Gaussian process reconstructions of the Hubble diagram \citep{2022MNRAS.514.5827S}, and joint analysis with various observations \citep{2023PDU....4201369O,2024PDU....4601558O,2024JCAP...06..051P,2024arXiv241209409O}. Further, in \citep{2017PhRvD..95j4034L}, the authors analyze the polarizations of linear gravitational waves in $f(R)$ gravity, demonstrating that $f(R)$ gravity propagates three polarization modes, which is different from the two polarization modes predicted by GR. Recently, the secondary gravitational waves (scalar induced gravitational waves) in $f(R)$ gravity was also studied \citep{2024JCAP...12..021Z}. Scalar induced gravitational waves originate from nonlinear interactions in the early universe, which provides a promising avenue for probing $f(R)$ gravity in the early universe. Furthermore, as a dark energy model, $f(R)$ gravity with curvature-matter interaction is also studied via the dynamical analysis approach \citep{2024arXiv241220209C}. This progression has established $f(R)$ gravity as one of the most rigorously tested modified gravity frameworks, though several theoretically viable models remain observationally unconstrained. 

In the present study, we consider three widely discussed $f(R)$ models:\\
(i) {\bf Hu-Sawicki model} \citep{2007PhRvD..76f4004H}: Designed to mimic $\Lambda$CDM at high redshift while introducing deviations at late times.\\
(ii) {\bf Starobinsky model} \citep{1980PhLB...91...99S}: Originally introduced as an inflationary model, it also serves as a modified gravity model for the late-time acceleration.\\
(iii) {\bf ArcTanh model} \citep{2018PhRvD..97b3525P}: A more recent parametrization that introduces a smooth transition from $\Lambda$CDM-like behavior to a modified gravity regime.\\

{These theories provide compelling phenomenological alternatives to dark energy for explaining the accelerated expansion of the universe. Their appeal lies in their ontological simplicity, proposing that cosmic acceleration emerges naturally from the dynamics of spacetime itself, without invoking exotic dark energy components.}

{ To constrain these models rigorously, we utilize a joint analysis of both early- and late-time cosmological datasets:}\\
(a) Cosmic Chronometers (Hubble parameter measurements) \citep{2002ApJ...573...37J}, (b) Type Ia Supernovae (Union 3.0 sample) \cite{2023arXiv231112098R}, (c) Baryon Acoustic Oscillations (BAO) from DESI-2025 \citep{2025arXiv250314743D,2025arXiv250314738D}, (d) Gamma-Ray Bursts (GRBs) \citep{2021JCAP...09..042K}, and Cosmic Microwave Background (CMB) \cite{2020A&A...641A...6P}.

{This joint analysis breaks  degeneracies between modified gravity parameters, particularly at high redshifts (CMB), and provides significantly tighter constraints compared to analyses using only low-redshift data. The multi-probe approach tests unified cosmic evolution from inflation to late-time acceleration.

Thus, our analysis provides independent} constraints on the expansion history of the universe, allowing us to test whether $f(R)$ models can serve as viable alternatives to the standard $\Lambda$CDM model. Our approach follows a Bayesian statistical framework using the Markov chain Monte Carlo (MCMC) method to obtain the best-fit values of the model parameters.\\

This paper is organized as follows: In Section II, we describe the theoretical background of $f(R)$ gravity and introduce the three chosen models. Section III outlines the dataset and methodology used for parameter estimation. In Section IV, we present the constraints obtained from different observational datasets. Section V discusses the implications of our results, and Section VI concludes with a summary of our findings.


\section{Mathematical Setup}
In this section we briefly review the main theoretical framework of the $f(R)$ models.\\

The modified Einstein-Hilbert action of gravity is given by:

\begin{equation}\label{equ_action}
    {S}=\dfrac{1}{2\kappa^2}\displaystyle\int d^4x \sqrt{-g}f(R)+ S^{(m)},
\end{equation}
where $f(R)$ is some function of the Ricci Scalar $R$ and $\kappa^2=8\pi G$. 

The matter contribution is govern by the term $S^{(m)}$. By varying the action with respect the metric, we obtain the following field equations
\begin{equation}\label{equ_action_equation}           
    f^\prime(R)R_{\mu\nu}-\dfrac{1}{2}f(R)g_{\mu\nu}-\left[\nabla_\mu\nabla_\nu-g_{\mu\nu}\Box\right]f^\prime(R)=\kappa^2 T_{\mu\nu},
\end{equation}
where $f^\prime(R)=\partial f(R)/\partial R$ and $\Box=g^{ab}\nabla_a\nabla_b$. Further, the $T_{\mu\nu}$ is the energy-momentum tensor corresponding to matter part of the above action. 

For an expanding, homogeneous and isotropic universe, the general spatially flat Friedmann-Lema$\hat{i}$tre-Robertson-Walker (FLRW) metric is defined as
\begin{equation}\label{equ_flrw}
    ds^2=-dt^2+a^2(t)\left[dr^2+r^2\left(d\theta^2+\sin^2(\theta)d\phi^2\right)\right],
\end{equation}
where $a(t)=1/\left(1+z\right)$ is scale factor and $z$ is the cosmological redshift.

Further, for an expanding universe, we consider a perfect fluid and corresponding to this, the energy-momentum tensor is $T^{\mu\nu}=(p+\rho)u^\mu u^\nu+pg^{\mu\nu}$, where, $u^a$ is the four-velocity of the fluid such that $u^au_a=-1$. The terms $p$ and $\rho$ are the total pressure and total energy density of the fluid respectively and each term has the matter and radiation components such that $p=p_r+p_m$ and $\rho=\rho_r+\rho_m$, where matter pressure is $p_m = 0$ and density of radiation is $p_r = \rho_r/3$. By assuming no interaction between matter and radiation, and considering the validity of the Bianchi identity for individual components, we obtain $\nabla^\mu T_{\mu\nu}=0$. This condition leads to the conservation of energy-momentum for both matter and radiation. Therefore, we can write:
\begin{equation}\label{equ_conservation_1} 
    \dot{\rho}_m + 3H\rho_m = 0, 
\end{equation} 

\begin{equation}\label{equ_conservation_2} 
    \dot{\rho}_r + 4H\rho_r = 0, 
\end{equation}
where over-dot $(\dot~)$ denotes a derivative with respect to the cosmic time $(t)$. The energy density of matter and radiation are given as $\rho_r=\rho_{r0}a^{-4}$ and $\rho_m=\rho_{m0}a^{-3}$. The Hubble parameter, $H(t)$ is defined as  $H(t)=\dot a(t)/a(t)$. 

The components of Einstein tensor are given as      
\begin{equation}\label{equ_einstein}
    G^0_{~0}=-3H^2, ~~~~~~~~~~~~~~~~G^i_{~i}=-\left(2\dot{H}+3H^2\right).
\end{equation}

Now, using \eqref{equ_action_equation} and  \eqref{equ_einstein}, we derive the modified Friedmann equations~\cite{2012PhR...513....1C,2010LRR....13....3D} as 
\begin{align}\label{equ_friedmann}
    H^2=&\dfrac{1}{3f^\prime(R)}\Bigg[8\pi G\left(\rho_r+\rho_m\right)-\frac{1}{2}\left(f(R)-Rf^\prime(R)\right) \notag \\ 
    & \hspace{3cm}\quad -3H\dot{f}^\prime(R) \Bigg], \\
    \dot H=& -\dfrac{1}{2f^\prime(R)}\Bigg[8\pi G\left(\rho_m+\frac{4}{3}\rho_r\right)+\Ddot{f}^\prime(R) \notag \\  
    &\hspace{3cm} \quad -H\dot f^\prime(R) \Bigg].
    \end{align}

Further, the contraction of Ricci tensor gives the Ricci scalar as  
\begin{equation}\label{equ_ricci_scalar}
    R=6\left(2H^2+\dot{H}\right).    
\end{equation}

Now, for example, if we consider $f(R)=R$ in the above metric formulism, then the field equations given in \eqref{equ_action_equation} leads to the nominal Einstein's equations corresponding to the Einstein de Sitter model. On the other hand, by setting $f(R)=R-2\Lambda$, one can recovers the one of the well-established model, the cosmological constant cold dark matter $(\Lambda \text{CDM})$ model. 

We would like to stress that the selection of the function $f(R)$ should consistent with both current cosmological \cite{2001PhLB..506...13H,2007PhRvD..75d4004S} and solar system observations~\cite{2007PhRvD..76f4004H}. In addition, the choice of $f(R)$ models should satisfy some strong conditions: $f^\prime(R)>0$ for $R\geq R_0>0$, where $R_0$ is Ricci scalar at current epoch. We also require the following conditions if the final attractor is a de Sitter spacetime with Ricci scalar $R_1$:  (i). $f^\prime(R) > 0$ for $R\geq R_1 > 0$, (ii). $f^{\prime\prime}(R)>0$ for $R\geq R_0>0$, (iii). $f(R)\approx R-2\Lambda$ for $R>>R_0$, and (iv). $0<\frac{Rf^{\prime\prime}(R)}{f^\prime(R)}(r)<1$ at $r=\frac{Rf^\prime(R)}{f(R)}=-2$. 

In this analysis, we explore the following three viable models and constrain them with the recent cosmological observations.

\subsection{Hu-Sawicki model}

{The Hu-Sawicki (HS) model~\cite{2007PhRvD..76f4004H} is a theoretically well-known $f(R)$ gravity framework designed to explain cosmic acceleration while satisfying local gravity constraints. Its defining functional form is:
\begin{equation}\label{equ_hs_model}
    f(R) = R - m^2 \dfrac{c_1 \left(\dfrac{R}{m^2}\right)^n}{1 + c_2 \left(\dfrac{R}{m^2}\right)^n},
\end{equation}
where $m^2 \equiv \Omega_{m0} H_0^2$ sets the curvature scale (comparable to the present Ricci scalar $R_0$), with $\Omega_{m0}$ and $H_0$ being the current matter density and Hubble parameter respectively. The dimensionless parameters $c_1$, $c_2$ and positive constant $n$ control the deviation from General Relativity.

The model achieves its physical relevance through two key features. First, its chameleon screening mechanism~\cite{2007JCAP...02..022N,2007PhRvD..76f3505F,2007PhRvL..98m1302A} suppresses modifications in high-density regions (e.g., Solar System) while allowing cosmic acceleration at cosmological scales. Second, with $c_1/c_2 \propto \Lambda_{\mathrm{obs}}/m^2$, it exactly reproduces $\Lambda$CDM expansion history, offering a viable dark energy alternative. 

These advantages come with theoretical challenges. The nonlinear curvature dependence ($n \geq 1$) may lead to stability issues and singularities in strong-gravity regimes. Nevertheless, the HS model remains a benchmark for modified gravity due to its observational consistency.

The model can be reexpressed in $\Lambda$CDM form through the transformation~\cite{2013CQGra..30a5008B}:
\begin{equation}\label{equ_hs_lcdm}
    f(R) = R - 2\Lambda\left(1 - \dfrac{1}{1 + \left(\dfrac{R}{b\Lambda}\right)^n}\right),
\end{equation}
where $\Lambda = m^2c_1/(2c_2)$ and $b = 2c_2^{1-1/n}/c_1$. This representation clarifies its connection to standard cosmology.}

\subsection{Starobinsky model}

{The Starobinsky model~\cite{1980PhLB...91...99S} represents a foundational $f(R)$ gravity framework originating from inflationary cosmology. Its original form,
\begin{equation}\label{equ_staro_original}
    f(R) = R + \frac{R^2}{6M^2},
\end{equation}
where $M$ is the inflation mass scale, produces a scalar degree of freedom (scalaron) that generates a plateau-like potential. This potential drives slow-roll inflation consistent with Planck CMB observations~\cite{2020A&A...641A...6P}, which makes it one of the most successful inflationary scenarios \footnote{{The recent observational data from the Atacama Cosmology Telescope (ACT) \cite{ACT:2025fju,ACT:2025tim} indicates a higher value of power spectrum index $n_s$ relative to Planck, which appears to disfavor the Starobinsky inflation mode \eqref{equ_staro_original}. In this paper, we only consider the following generalized form of Starobinsky mode \eqref{equ_staro_model}, which is responsible for the late-time cosmic acceleration.}}

The generalized form of the model is given by:
\begin{equation}\label{equ_staro_model}
    f(R) = R - c_1m^2\left(1 - \left(1 + \frac{R^2}{m^4}\right)^{-n}\right),
\end{equation}
where the constants $c_1$, $m^2 \equiv \Omega_{m0}H_0^2$, and $n$ maintain the same definitions as in the Hu-Sawicki case. This extension preserves the UV-complete nature of the original model while introducing additional flexibility for late-time cosmology applications.

Similar to the HS model, the Starobinsky formulation can be expressed in a $\Lambda$CDM-like form:
\begin{equation}\label{equ_staro_lcdm}
    f(R) = R - 2\Lambda\left(1 - \frac{1}{\left(1 + \left(\frac{R}{b\Lambda}\right)^2\right)^n}\right),
\end{equation}
where $\Lambda = c_1m^2/2$ and $b = 2/c_1$. This representation explicitly connects the inflationary model to standard cosmology.

The Starobinsky model remains particularly significant for several reasons: (1) its original $R^2$ form avoids cosmological singularities, (2) it provides excellent agreement with CMB measurements, and (3) its generalized versions maintain these advantages while extending applicability to dark energy phenomenology. These features collectively establish it as a cornerstone of modified gravity theories.}




\subsection{ArcTanh model} 
{The ArcTanh model, introduced by P\'{e}rez Romero and Nesseris~\cite{2018PhRvD..97b3525P}, offers a distinctive approach to modified gravity through its non-polynomial formulation. The model is defined by
\begin{equation}\label{equ_arct_model}
    f(R) = R - 2\Lambda\left(\dfrac{1}{1 + b\,\text{arctanh}\left(\dfrac{\Lambda}{R}\right)}\right),
\end{equation}
where $\Lambda$ corresponds to the cosmological constant and $b$ is a dimensionless coupling parameter. The hyperbolic arctangent function ensures bounded nonlinear behavior that prevents curvature singularities while naturally incorporating a screening mechanism at high curvatures ($R \gg \Lambda$). In the low-curvature limit ($R \to \Lambda$), the model smoothly recovers standard $\Lambda$CDM cosmology.

The ArcTanh model demonstrates improved behavior in strong-gravity regimes compared to polynomial $f(R)$ theories, with a stable transition between cosmological and astrophysical scales. Although less widely studied than the Hu-Sawicki or Starobinsky models, its unique mathematical structure provides valuable insights into gravity's behavior under extreme conditions and represents a significant addition to the landscape of modified gravity theories.} 


\vspace{6mm}
In this paper, we systematically explore the functional space of viable $f(R)$ theories that remain small perturbations around the $\Lambda$CDM model. Without loss of generality, we consider $f(R)$ models of the form:
\begin{equation}
    f(R)=R-2\Lambda h(R,b),
\end{equation}
where
\begin{equation}
    h(R,b)=
\begin{cases}
1-\dfrac{1}{1+\left(\dfrac{R}{b\Lambda}\right)^n}                &  (\text{for Hu-Sawichi model}),  \vspace{1mm}\\
1-\dfrac{1}{\left(1+\left(\dfrac{R}{b\Lambda}\right)^2\right)^n} &  (\text{for Starobinsky model}),  \vspace{1mm}\\
\dfrac{1}{1+b~\text{arctanh}{\left(\dfrac{\Lambda}{R}\right)}}   &  (\text{for ArcTanh model}).  \\
\end{cases}
\end{equation} 

Due to the known degeneracy between the parameter $n$ and the present matter density $\Omega_{m0}$, which prevents their simultaneous constraint from observational data alone, we fix $n = 1$ in our analysis. This choice is motivated by its consistency with theoretical expectations and observational constraints, as this value is widely adopted in the literature~\cite{2013PhRvD..87l3529B,2009PhRvD..80h3505S,2018PhRvD..97b3525P,2017JCAP...01..005N} and provides a better fit to observations compared to larger values of $n$. \\

In fact, one can understands that $\Lambda$CDM model is the limiting case of all these three model as 
\begin{align}
    \lim_{b\rightarrow0} h(R,b)&=1\rightarrow f(R)=R-2\Lambda, \\
    \lim_{b\rightarrow\infty}h(R,b)&= 0\rightarrow f(R)=R. 
\end{align}

To obtain the approximate theoretical Hubble parameter expression $(H_{\mathrm{th}}(z))$ in terms of redshift $(z)$ for these models, we adopt the approach followed by Basilakos et al.~\cite{2013PhRvD..87l3529B} and solve the modified Friedmann equation (\eqref{equ_friedmann}). For more details, please refer to Sultana et al.~\cite{2022MNRAS.514.5827S}.


\section{Dataset and Methodology}\label{sec_data_method}

In this analysis, we use four recent observational datasets of Hubble parameter measurements (Cosmic Chronometers), Type Ia supernovae (Union 3.0), Baryon Acoustic Oscillations (DESI~2025), and Gamma Ray Bursts (GRBs). Further, to optimize the above mentioned models, we adopt a Bayesian statistics based Python module \textbf{ {\texttt{emcee}}}\footnote{\url{https://emcee.readthedocs.io/en/stable/}} and find the same best fit values of parameters.

\subsection{Hubble Parameter (Cosmic Chronometers)}
A direct and model-independent way to constrain the expansion history of the universe is provided by the ages of the oldest objects in the universe at high redshift. This method is known as ``Cosmic Chronometers'' and was first proposed by Jimenez and Loeb in 2002 \cite{2002ApJ...573...37J}. Using this method, the Hubble parameter, $H(z)$, is expressed as
\begin{equation}
    H(z)=-\dfrac{1}{\left(1+z\right)}\dfrac{dz}{dt}.
\end{equation}

Thus, to measure the Hubble parameter for a given redshift, one requires the redshift and its derivative with respect to cosmic time accurately. For astrophysical objects, we can directly measure the redshift with negligible uncertainty. The one factor that remains unknown, is the differential age of galaxies, $dt$. The stars, in young evolving galaxies, are continually born in early developing galaxies and the emission spectra will be dominated by the young stellar population.  Hence to estimate accurately the differential aging of the universe, passively evolving red galaxies are used as their light is mostly dominated by the old stellar population \cite{2002ApJ...573...37J}. Thus by estimation of the differential age evolution of the universe in a given interval of the redshift from various methods like full spectrum fitting, absorption feature analysis and calibration of specific spectroscopic features~\cite{1994ApJS...95..107W,2011MNRAS.412.2183Tf, 2012JCAP...08..006M}, we use 32 data points of Hubble parameter {measurements as well as its covariance matrix given in analysis of \cite{2022LRR....25....6M}}.

\subsection{Type Ia Supernovae (Union 3.0)}

In this analysis, we use the most recent compilation of observations of the Type Ia supernovae named `Union 3.0'. This sample is largest datasets of 2087 SNIa in the redshift range $0.01<z<2.26$ which are obtained from 24 datasets after analyzed through an unified Bayesian framework in the redshift range  \cite{2023arXiv231112098R}. We should caution that currently the binned distance moduli for this data sample is available, therefore, we use only the binned datapoints in our analysis. 

In the observed dataset, we have an observations of apparent magnitude $(m)$. Once we know the value of absolute magnitude, the distance modulus $(\mu)$ can be expressed in terms of luminosity distance $(d_L)$ as
\begin{equation}\label{equ_distance_modulus}
    \mu(z)=m_B(z)-M_B=5 \log _{10}\left[\frac{d_L(z)}{\mathrm{Mpc}}\right]+25,
\end{equation}
where $d_L$, in case of flat FLRW metric is 
\begin{equation}\label{eq_dl_h}
    d_L(z)=c\left(1+z\right)\displaystyle\int_0^z \dfrac{dz^\prime}{H(z^\prime)}.
\end{equation}

Recent research suggests that the luminosity (or absolute magnitude) of Type Ia supernovae does not evolve with redshift~\cite{2019A&A...625A..15T,2022JCAP...01..053K}. So, it is generally accepted that the Type Ia supernovae sample is normally distributed with a mean absolute magnitude of $M_B = -19.22$~\cite{2020PhRvD.101j3517B}.

\subsection{Baryon Acoustic Oscillations (DESI-2025)}           

Another valuable cosmological probe comes from the baryon acoustic oscillation (BAO) data. In this work, {we use the second-year BAO measurements from DESI, which cover three years of observations \citep{2025arXiv250314743D,2025arXiv250314738D}. The analysis uses the full covariance matrix to account for correlations between different distance indicators and provides a consistent statistical treatment.} 


The observed quantities in these measurements are given as follows:
\begin{align} 
    \text{First quantity}: \quad & \dfrac{d_M(z)}{r_d} = \dfrac{d_L(z)}{r_d\left(1+z\right)},  \label{eq_dM} \\
    \text{Second quantity}: \quad & \dfrac{d_H(z)}{r_d} = \dfrac{c}{r_d H(z)}, \label{eq_dH} \\
    \text{Third quantity}: \quad & \dfrac{d_V(z)}{r_d} = \dfrac{\left[z d_H(z) d_M^2(z)\right]^{1/3}}{r_d}, \label{eq_dV}
\end{align}
where $d_M(z)$ is the transverse co-moving distance, $d_H(z)$ is the Hubble distance, $H(z)$ is the Hubble parameter, $d_V(z)$ represents the volume-average angular diameter distance and the $r_d$ denotes the sound horizon at the drag epoch, which depends on the matter and baryon physical energy densities and the effective number of extra-relativistic degrees of freedom.    

In the BAO analysis, it is notice that there is strong degeneracy between $H_0$ and $r_d$. Therefore, to break this degeneracy, we adopt a prior value of baryon density parameter $\Omega_{b0}=0.02218\pm0.00055$, which is used to compute $r_d$ in our analysis~\cite{2025JCAP...02..021A}.

\subsection{Gamma Ray Bursts (GRBs)}

For this analysis, we use a linear regression relation using the logarithms of $E_{\mathrm{iso}}$ and $E_{\mathrm{P}}$. This relation is generally referred to as the Amati relation which is basically a correlation between isotropic equivalent energy ($E_{\mathrm{iso}}$) and spectrum peak energy in the comoving frame ($E_{\mathrm{P}}$). The Amati relation can be parametrized as 
\begin{equation}\label{GRB_1}
    \log\left[\dfrac{E_{\mathrm P}}{1~\mathrm{keV}}\right]=m\log\left[\dfrac{E_{\mathrm{iso}}}{1~\mathrm{erg}}\right]+c,
\end{equation}
where $m$ (slope) and $c$ (intercept) are the two parameters of the Amati correlation. 

The isotropic equivalent energy $E_{\mathrm{iso}}$ is given by
\begin{equation}\label{GRB_2}
    E_{\mathrm{iso}}=\dfrac{4\pi d_L^2S_{\mathrm{bolo}}}{\left(1+z\right)}.
\end{equation} 

Here $S_{\text{bolo}}$ is the bolometric fluence and $d_L$ is luminosity distance which is related to the comoving distance as $d_L = d_{C}(1+z)$. The factor $(1+z)$, in \eqref{GRB_2}, accounts for the cosmological time dilation effect. The peak energy in the comoving frame, $E_P$, is related to the observed peak energy $E_{P}^\mathrm{obs}$ by
\begin{equation}\label{GRB_3}
    E_\mathrm{P}=E_{\mathrm{P}}^{\text{obs}}\left(1+z\right),
\end{equation}

We use GRB data having $220$ data points (referred to as A220) in the redshift range, $0.0331 \leq z \leq 8.20$. This data is consolidated in literature (Tables 7 and 8 of Ref.~\cite{2021JCAP...09..042K}). In the data, corresponding to each sample source, the name of the GRB, its redshift, spectral peak energy in the rest frame ($E_p$), and measurement of the bolometric fluence ($S_{bolo}$) along with $1\sigma$ confidence level are mentioned. A220 is the union of two samples,  A118 ($0.3399 \leq z \leq 8.2$) and A102 ($0.0331 \leq z \leq 6.32$). The A118 data that has 118 long GRBs is further composed of two subsamples, i.e., 93 GRBs and 25 GRBs, collected from~\cite{2016A&A...585A..68W} and \cite{2019ApJ...887...13F}, respectively. A102 consist of 102 long GRBs taken from~\cite{2016A&A...585A..68W, 2019MNRAS.486L..46A}.  

We note that in the literature, the best-fit values of the Amati correlation parameters have been found to be constant and same for all cosmological models~\cite{2021JCAP...09..042K,2023arXiv230716467X}. This suggests that the Amati relation is independent of different cosmological models and that the GRBs within a given data set are standardized. Therefore, in this work, we fix the Amati correlation parameters based on Ref.~\cite{2023JCAP...07..021K}. Since our aim is to test modified cosmological models, we do not expect variations in the Amati parameters to significantly impact our results.

\subsection{Cosmic Microwave Background (CMB)}

{The temperature and polarization data from the CMB provide strong constraints on cosmological models. To use the full CMB dataset, one must solve the modified Boltzmann equations specific to the underlying theory of gravity. These equations depend on the cosmological perturbation theory and differ significantly from those used in standard Einstein gravity. Numerical codes such as \texttt{CAMB}~\cite{2000ApJ...538..473L} and \texttt{CLASS}~\cite{2011JCAP...07..034B} are based on the standard framework and cannot be directly applied to alternative theories. Solving the full set of equations for modified gravity models demands significant computational resources.

Recent work by Chen et al.~\cite{2019JCAP...02..028C} has compared constraints from the full CMB power spectrum with those obtained using CMB distance priors. Their study shows excellent agreement between the two methods across several dark energy models. This result supports the use of distance priors as a reliable alternative when working with non-standard cosmological models.

In the present analysis, we adopt the CMB distance priors provided in~\cite{2019JCAP...02..028C}, derived from the Planck 2018 temperature and polarization data (TT, TE, EE + lowE). These priors encode the key geometric information from the full CMB spectrum in terms of the shift parameter $R$, the acoustic scale $l_A$, and the physical baryon density $\Omega_b h^2$. They allow us to include CMB constraints without computing the full set of perturbations.

The CMB distance prior dataset includes the shift parameter $R$, the acoustic scale $l_A$, and the baryon density parameter $\Omega_b h^2$. These quantities are calculated as
\begin{equation}
l_A = (1 + z_*) \frac{\pi D_A(z_*)}{r_s(z_*)}, \quad R = \frac{(1 + z_*) D_A(z_*) \sqrt{\Omega_m H_0^2}}{c},
\end{equation}
where $z_*$ is the redshift at the photon decoupling epoch and $r_s(z_*)$ is the comoving sound horizon at that epoch. The comoving sound horizon is given by
\begin{equation}\label{rs_cmb_eq}
    r_s(z)=\frac{c}{H_0} \int_0^{1 /(1+z)} \frac{d a}{a^2 E(a) \sqrt{3\left(1+\frac{3 \Omega_b h^2}{4 \Omega_\gamma h^2} a\right)}},
\end{equation}
where the photon density term is
$$
\frac{3}{4 \Omega_\gamma h^2}=31500\left(\frac{T_{\mathrm{CMB}}}{2.7\, \mathrm{K}}\right)^{-4}, \quad T_{\mathrm{CMB}}=2.7255\, \mathrm{K}~\cite{2009ApJ...707..916F}.
$$

While computing Eq.~\eqref{rs_cmb_eq}, we include the radiation contribution, i.e., $\Omega_r=\Omega_{r0}a^{-4}$, in the expression of the dimensionless Hubble parameter $E(a)$. Here, the present radiation density is expressed as $\Omega_{r0}=\dfrac{\Omega_{m0}}{1+z_{\mathrm{eq}}}$. In this analysis, we adopt the values of the matter-radiation equality redshift $z_{\mathrm{eq}}$, the photon decoupling redshift $z_*$ and $\Omega_{b} h^2$ from the Planck 2018 survey~\cite{2020A&A...641A...6P}.

{
It should be noted that the compressed CMB likelihood \cite{2019JCAP...02..028C} inherits $\Lambda$CDM dependencies from its construction. Following the approach established in previous studies \cite{2008JCAP...09..008C,2024PDU....4601641S,2024PDU....4601668H,2024arXiv241209064L,2025arXiv250308236R}, we combine this CMB analysis with late-time cosmological observations (DESI BAO and Pantheon+ supernovae) to obtain balanced constraints. While the compressed data primarily tests consistency with $\Lambda$CDM background geometry, our core conclusions rely on the synergy of multiple probes, which together provide robust tests of modified gravity scenarios.
}
}

\subsection{Methodology}

In this analysis, we define an individual chi-square corresponding to each observational datasets. \\

For the Cosmic Chronometer dataset, we use the following Chi-square as     
{
\begin{equation}\label{eq_chi_1}
\begin{split}
\chi^{2}_{\mathrm{CC}} = & \left[\mathrm{{H}_{th}}\left(z_i;H_0,\Omega_{m0},b\right)-\mathrm{{H}_{obs}}\left(z_{i}\right)\right]^T \\
&  \operatorname{{Cov}_{ij}}^{-1} \left[\mathrm{{H}_{th}}\left(z_j;H_0,\Omega_{m0},b\right)-\mathrm{{H}_{obs}}(z_j)\right].
\end{split}
\end{equation}
where 
$H_{th}$ and $H_{obs}$ are the theoretical and observed Hubble parameters, respectively. The theoretical Hubble parameter $H_{th}$ is derived from modified gravity models, such as the Hu-Sawicki model, the Starobinsky model, and the Arctanh model. The $\operatorname{Cov}_{i j}$ represents the total uncertainty in the Hubble parameter which includes the effect of both statistical as well as systematic errors. Thus in our analysis, we consider the full covariance matrix $\operatorname{Cov}_{i j}$ as 
\begin{equation}\label{cov_1}
    \operatorname{Cov}_{i j}=\operatorname{Cov}_{i j}^{\text {stat }}+\operatorname{Cov}_{i j}^{\mathrm{syst}},
\end{equation}
where $\operatorname{Cov}_{i j}^{\mathrm{stat}}$  denote the contributions to the covariance due to statistical errors. 

The contribution due to systematic errors is given by $\operatorname{Cov}_{i j}^{\mathrm{syst}}$. To compute the systematic errors carefully we refer to the analysis carried out by Moresco et al.~\cite{2022LRR....25....6M} where, we have considered the four main contributions in the covariance matrix, i.e. initial mass function, stellar library, metallicity, and stellar population synthesis models~\cite{2020ApJ...898...82M}. For a detailed discussion of the full covariance matrix please refer to Section 3.1.4 of Moresco et al.~\cite{2022LRR....25....6M}. 

For the Union 3.0, the Chi-square takes the form as
\begin{equation}\label{chi2_Pan}
    \chi^2_{\mathrm{Union~3.0}} = \Delta \mu^T \cdot C^{-1} \cdot \Delta \mu,
\end{equation}
where \( C = D_{\text{stat}} + C_{\text{sys}} \) is the total covariance matrix. Here, \( D_{\text{stat}} \) is the diagonal covariance matrix of statistical uncertainties, and \( C_{\text{sys}} \) represents the systematic covariance matrix. The vector 
\begin{equation}
    \Delta \mu = \mu_{\text{obs}}(z_i) - \mu_{\text{th}}\left(z_i;H_0,\Omega_{m0},b\right),
\end{equation}
which is given in \eqref{equ_distance_modulus} and  \eqref{eq_dl_h}.\\

The Chi-square for the BAO dataset is 

\begin{equation}
\begin{split}
    \chi_{\mathrm{BAO}}^2=& \left[{X_{\mathrm{th}}\left(z_i;H_0,\Omega_{m0},b\right)-X_{obs}\left(z_i\right)}\right]^T.\operatorname{Cov}_{i j}.\\ & \hspace{1.2cm}\left[{X_{\mathrm{th}}\left(z_j;H_0,\Omega_{m0},b\right)-X_{obs}\left(z_j\right)}\right]
\end{split}
\end{equation}
where, the term $X$ takes the quantities from  \eqref{eq_dM}, \eqref{eq_dH} and  \eqref{eq_dV}.     
     
For the GRBs dataset, the Chi-square is given as
\begin{equation}\label{GRB_7}
    \chi^2_{GRBs}=\sum_i \frac{\left[\mu_{\mathrm{th}}\left(z_i;H_0,\Omega_{m0},b\right)-\mu_{obs}\left(z_i\right)\right]^2}{\sigma_{\mu_i}^2},
\end{equation}
where $\sigma_\mu^2(z_i)$ is the total uncertainty, which accounts for the intrinsic scatter of GRBs ($\sigma_{\mathrm{int}}$) and the observational errors in $S_{\mathrm{bolo}}$ and $E_{\mathrm{iso}}$. \\

{In CMB distance priors, we use the values $R$, $l_A$, and $\Omega_b h^2$, along with their correlation matrix to build the full covariance matrix $C_{ij}$. The corresponding chi-square expression is given by
\begin{equation}
\chi^2_{\text{CMB}} = \Delta V(\Theta)^T C_{\text{CMB}}^{-1} \Delta V(\Theta),
\end{equation}
where $\Delta V = V(\Theta) - V$ and $V(\Theta) = \{R(\Theta), l_A(\Theta), \Omega_b h^2\}$.
}

To constrain the cosmological parameters, we employ Bayesian statistics using the Python module \textbf{{\texttt{emcee}}}, which is an affine-invariant Markov chain Monte Carlo (MCMC) sampler. We maximize the total likelihood function, which follows the relation:  
\begin{equation}
    \mathcal{L} \sim \exp \left(-\frac{\chi_T^2}{2} \right),
\end{equation}  
where the total chi-square function, $\chi_T^2$, is given by the sum of the chi-square values from different observational data sets:  
\begin{equation}
    \chi^2_T = \chi^2_{CC} + \chi^2_{BAO} + \chi^2_{Union 3.0} + \chi^2_{GRBs} + \chi^2_{CMB}.
\end{equation}  

Since these observational datasets are independent, their corresponding chi-square contributions can be added directly to obtain the total chi-square, $\chi_T^2$. This approach allows us to perform a joint analysis and derive constraints on the cosmological parameters in a statistically consistent manner.

{In this work, we adopt flat priors for all parameters (Table \ref{tab_prior}). Our MCMC analysis uses 10 walkers with 10,000 steps each to explore the parameter space. We discard the first 20\% of steps as burn-in and analyze the posterior distributions using the remaining samples. To verify chain convergence, we perform an auto-correlation analysis by computing the integrated auto-correlation time $\tau_f$ using the \textbf{\textit{autocorr.integrated\_time}} function of the \textbf{\textit{emcee}} package. For more details, please see ref.\cite{2013PASP..125..306F} .

\begin{table}[h]
	\centering
	\renewcommand{\arraystretch}{2}
	\begin{tabular}[b]{| l | l|}\hline
		Parameter & Prior Range\\ \hline \hline
		$H_0~[\mathrm{km\,s^{-1}\,Mpc^{-1}}]$ & $\mathbb{U}[0,100]$\\ \hline
		$\Omega_{m0}$ & $\mathbb{U}[0,1]$ \\ \hline
		$b$ & $\mathbb{U}[-1.5,1.5]$ \\ \hline
	\end{tabular}
	\caption{ The prior range of $H_0,~\Omega_{m0}$, and $b$.}
	\label{tab_prior}
\end{table}
    
}


\section{Results}\label{sec_results}

In this work, we constrain three $f(R)$ gravity models—the Hu-Sawicki model, Starobinsky model, and an $\mathrm{arctanh}$ model—using a joint analysis of recent cosmological observations. The datasets include Hubble parameter measurements (Cosmic Chronometers), Type Ia supernovae (Union 3.0), baryon acoustic oscillations (DESI~2025), Gamma-Ray Bursts (GRBs) and Cosmic Microwave Radiation (CMB). We account for full covariance structures and systematic uncertainties across all datasets to derive robust parameter constraints. 

\subsection*{Hu-Sawicki model}

The best-fit parameters for the Hu-Sawicki model are tabulated in Table~\ref{Hu_results}.     

\begin{table*}[!htbp] 
\centering
\renewcommand{\arraystretch}{1.5}
\setlength{\tabcolsep}{8pt}
\begin{tabular}{|l|c|c|c|c|}
\hline
\textbf{Parameter} & \textbf{CC} & \textbf{Union 3.0} & \textbf{DESI-2025} & \textbf{CC + Union 3.0 + DESI-2025 + GRBs + CMB} \\
\hline
$H_0~[\mathrm{km\,s^{-1}\,Mpc^{-1}}]$ & $66.074^{+5.057}_{-4.357}$ & $74.823^{+3.777}_{-3.159}$ & $69.310^{+0.469}_{-0.474}$ & $68.921^{+0.186}_{-0.209}$ \\
\hline
$\Omega_{m0}$ & $0.348^{+0.074}_{-0.074}$ & $0.293^{+0.067}_{-0.061}$ & $0.288^{+0.009}_{-0.008}$ & $0.299^{+0.002}_{-0.001}$ \\
\hline
$b$ & $0.002^{+0.708}_{-0.694}$ & $0.473^{+0.324}_{-0.478}$ & $0.295^{+0.168}_{-0.190}$ & $0.217^{+0.084}_{-0.080}$ \\
\hline
\end{tabular}
\caption{Best-fit values of model parameters at the 68\% confidence level for the Hu-Sawicki model from individual and joint analyses of observational datasets.}
\label{Hu_results}
\end{table*}


The $H_0$ shows notable variation across datasets: while Cosmic Chronometers (CC) produces a lower value\\ ($66.074^{+5.057}_{-4.357}~\mathrm{km\,s^{-1}\,Mpc^{-1}}$), Union 3.0 SNe favor a higher estimate ($74.823^{+3.777}_{-3.159}~\mathrm{km\,s^{-1}\,Mpc^{-1}}$), and DESI~2025 BAO data suggest an intermediate value ($69.310^{+0.469}_{-0.474}~\mathrm{km\,s^{-1}\,Mpc^{-1}}$). The combined analysis (CC + Union~3.0 + DESI-2025 + GRBs + CMB) converges to $H_0 = 68.921^{+0.186}_{-0.209}~\mathrm{km\,s^{-1}\,Mpc^{-1}}$ with reduce uncertainties compared to individual datasets. Similarly, the matter density parameter ($\Omega_{m0}$) tightens from $0.348^{+0.074}_{-0.074}$ (CC-only) to $0.299^{+0.002}_{-0.001}$ in the joint fit. The $b$ parameter constraints show all individual datasets—Cosmic Chronometers ($b = 0.002^{+0.708}_{-0.694}$), Union~3.0 SNe ($b = 0.473^{+0.324}_{-0.478}$), and DESI~2025 BAO ($b = 0.295^{+0.168}_{-0.190}$)—are consistent with $\Lambda$CDM ($b=0$) at the 95\% confidence level. However, the joint analysis of these probes gives $b = 0.217^{+0.084}_{-0.080}$ which represents a strong deviation from $\Lambda$CDM. This demonstrates that while no single dataset has sufficient statistical power to detect modified gravity effects, their combination reveals significant evidence for beyond-$\Lambda$CDM physics. 

The 1D marginalized posteriors and 2D joint confidence contours for the $H_0$, $\Omega_{m0}$, $b$ are shown in Fig.~\ref{Hu_contour_total}. 
The contours represent the $68\%$ and $95\%$ confidence levels. 





\begin{figure}[!htbp]    
    \centering
    \includegraphics[width=0.45\textwidth]{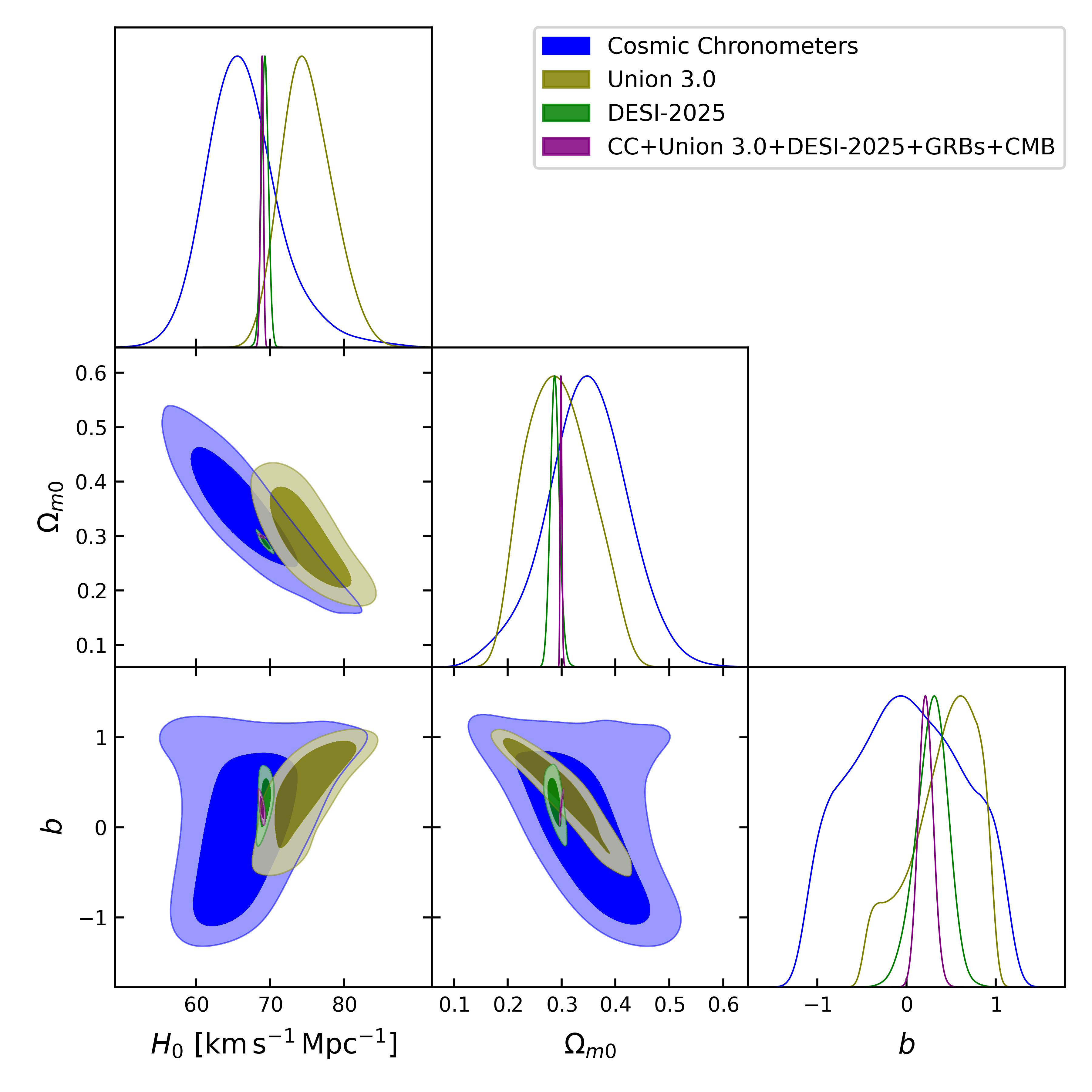}  
    \caption{68\% and 95\% confidence contours and posterior distributions of model parameters for the Hu-Sawicki model using individual and joint analyses of observational datasets. }
     \label{Hu_contour_total}
\end{figure} 


\subsection*{Starobinsky model}
The best-fit parameters for the Starobinsky model are summarized in Table~\ref{starobinsky_results}.
             
\begin{table*}[!htbp]
\centering
\renewcommand{\arraystretch}{1.5}
\setlength{\tabcolsep}{8pt}
\begin{tabular}{|l|c|c|c|c|}
\hline
\textbf{Parameter} & \textbf{CC} & \textbf{Union 3.0} & \textbf{DESI-2025} & \textbf{CC + Union 3.0 + DESI-2025 + GRBs + CMB} \\
\hline
$H_0~[\mathrm{km\,s^{-1}\,Mpc^{-1}}]$ & $66.226^{+4.043}_{-4.210}$ & $72.043^{+2.128}_{-2.148}$ & $66.715^{+1.404}_{-1.009}$ & $66.715^{+0.780}_{-0.777}$ \\
\hline
$\Omega_{m0}$ & $0.337^{+0.063}_{-0.049}$ & $0.333^{+0.028}_{-0.031}$ & $0.315^{+0.013}_{-0.013}$ & $0.318^{+0.007}_{-0.007}$ \\
\hline
$b$ & $-0.023^{+0.805}_{-0.770}$ & $0.935^{+0.195}_{-0.533}$ & $1.022^{+0.126}_{-0.289}$ & $1.040^{+0.119}_{-0.149}$ \\
\hline
\end{tabular}
\caption{Best-fit values of model parameters at the 68\% confidence level for the Starobinsky model from individual and joint analyses of observational datasets.}
\label{starobinsky_results}
\end{table*}

The $H_0$ values show variation across datasets: Cosmic Chronometers (CC) provide $66.226^{+4.043}_{-4.210}~\mathrm{km\,s^{-1}\,Mpc^{-1}}$, Union 3.0 SNe favor $72.043^{+2.128}_{-2.148}~\mathrm{km\,s^{-1}\,Mpc^{-1}}$, and DESI~2025 BAO data suggest $66.712^{+1.404}_{-1.009}~\mathrm{km\,s^{-1}\,Mpc^{-1}}$. The joint analysis (CC + Union~3.0 + DESI~2025 + GRBs + CMB) gives $H_0 = 65.885^{+0.988}_{-0.725}~\mathrm{km\,s^{-1}\,Mpc^{-1}}$ with reduced uncertainties compared to individual datasets. The matter density parameter ($\Omega_{m0}$) tightens from $0.337^{+0.063}_{-0.049}$ (CC-only) to $0.330^{+0.006}_{-0.009}$ in the joint fit. The best-fit value of the parameter $b$ indicates that DESI-2025 BAO ($b = 1.022^{+0.126}_{-0.289}$) favors stronger modifications compared to Union 3.0 SNe ($b = 0.935^{+0.195}_{-0.533}$). The joint analysis gives $b = 1.105^{+0.114}_{-0.180}$, which shows a significant deviation from $\Lambda$CDM.

The 1D posterior distributions and 2D joint confidence contours for the parameters $(H_0, \Omega_{m0}, b)$ are presented in Fig.~\ref{Staro_contour_total}.
with contours corresponding to the $68\%$ and $95\%$ confidence regions. Notably, the best fit value of parameter $b$ shows a weaker constraints in both the DESI~2025 and joint analyses
compared to other parameters.




\begin{figure}[!htbp]    
    \centering
    \includegraphics[width=0.45\textwidth]{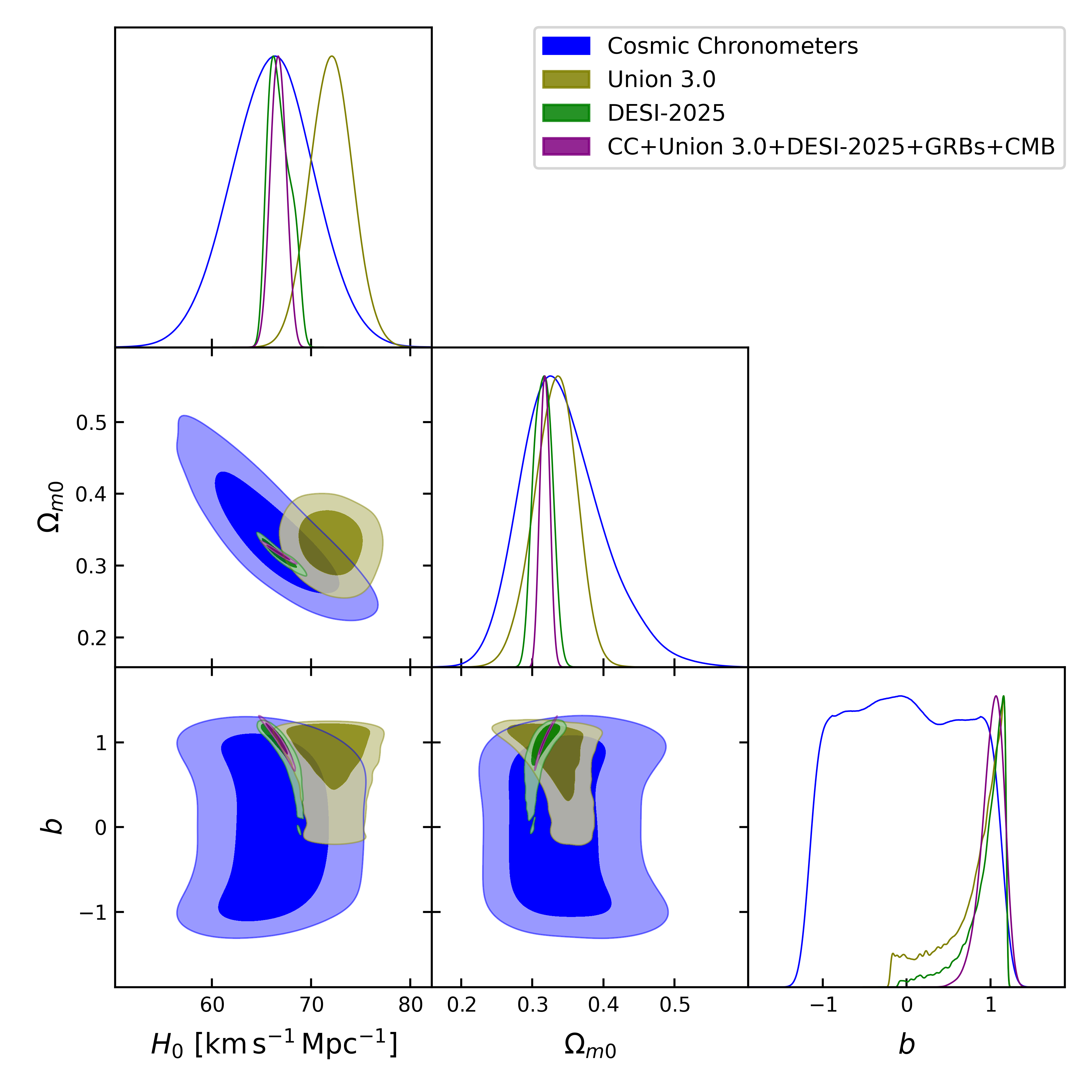}  
    \caption{68\% and 95\% confidence contours and posterior distributions of model parameters for the Starobinsky model using individual and joint analyses of observational datasets. }
     \label{Staro_contour_total}
\end{figure} 

\subsection*{ArcTanh model}
The best-fit parameters for the ArcTanh model are summarized in Table~\ref{ArcTanh_results}. 

\begin{table*}[!htbp]
\centering
\renewcommand{\arraystretch}{1.5}
\setlength{\tabcolsep}{8pt} 
\begin{tabular}{|l|c|c|c|c|}
\hline  
\textbf{Parameter} & \textbf{CC} & \textbf{Union 3.0} & \textbf{DESI-2025} & \textbf{CC + Union 3.0 + DESI-2025 + GRBs + CMB} \\ \hline
$H_0~[\mathrm{km\,s^{-1}\,Mpc^{-1}}]$ & $65.986^{+5.053}_{-4.541}$ & $74.635^{+3.726}_{-3.170}$ & $69.153^{+0.448}_{-0.492}$ & $68.849^{+0.205}_{-0.239}$ \\ \hline
$\Omega_{m0}$ & $0.351^{+0.078}_{-0.074}$ & $0.296^{+0.074}_{-0.065}$ & $0.289^{+0.008}_{-0.007}$ & $0.300^{+0.002}_{-0.002}$ \\ \hline
$b$ & $-0.007^{+0.724}_{-0.707}$ & $0.442^{+0.326}_{-0.525}$ & $0.338^{+0.179}_{-0.191}$ & $0.233^{+0.089}_{-0.086}$ \\ \hline
\end{tabular}
\caption{Best-fit values of model parameters at the 68\% confidence level for the Arctanh model from individual and joint analyses of observational datasets.}   

\label{ArcTanh_results}        
\end{table*}


The Hubble constant measurements range from \\ $65.986^{+5.053}_{-4.541}~\mathrm{km\,s^{-1}\,Mpc^{-1}}$ (CC) to $74.635^{+3.726}_{-3.170}~\mathrm{km\,s^{-1}\,Mpc^{-1}}$ (Union~3.0), with DESI~2025 BAO giving $69.153^{+0.448}_{-0.492}~\mathrm{km\,s^{-1}\,Mpc^{-1}}$. The joint analysis constrains $H_0 = 68.849^{+0.205}_{-0.239}~\mathrm{km\,s^{-1}\,Mpc^{-1}}$ and $\Omega_{m0} = 0.300^{+0.002}_{-0.002}$. For the modified gravity parameter, individual datasets show $b = -0.007^{+0.724}_{-0.707}$ (CC), $0.442^{+0.326}_{-0.525}$ (Union~3.0), and $0.338^{+0.179}_{-0.191}$ (DESI~2025), while the combined analysis gives $b = 0.233^{+0.089}_{-0.086}$, corresponding to a $2\sigma$ deviation from $\Lambda$CDM.

Fig.~\ref{Arct_contour_total} shows the constraints on the model parameters $(H_0, \Omega_{m0}, b)$ obtained from the individual and joint analyses of observational datasets. The contours represent the $68\%$ and $95\%$ confidence regions. These constraints highlight the bounds on parameters and the nature of their correlations within the ArcTanh model framework.




\begin{figure}[!htbp]    
    \centering
    \includegraphics[width=0.45\textwidth]{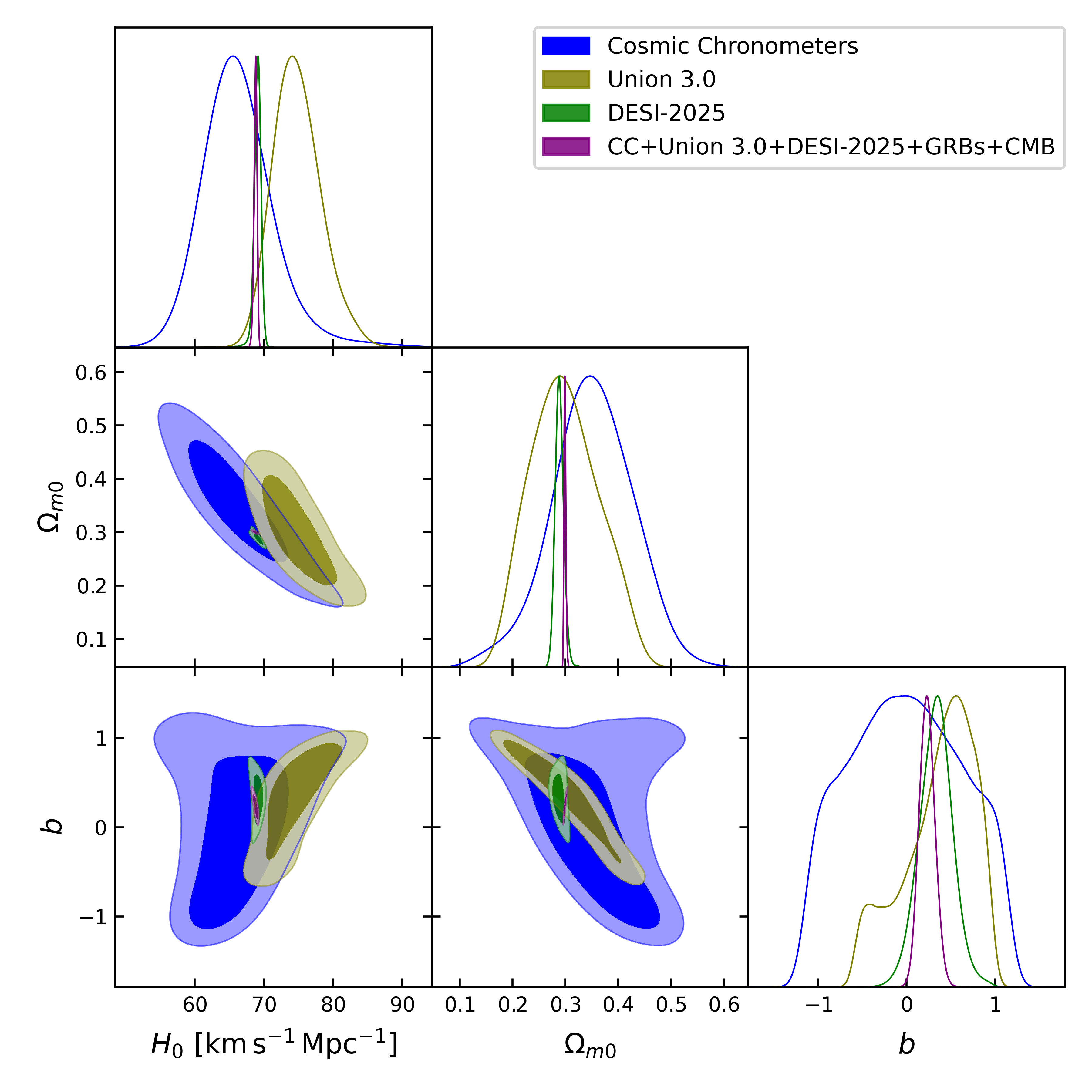}  
    \caption{68\% and 95\% confidence contours and posterior distributions of model parameters for the Arctanh model using individual and joint analyses of observational datasets.}
     \label{Arct_contour_total}
\end{figure} 


{
\subsection*{Model Comparison}

To evaluate and compare the performance of the cosmological models considered in this work, we use three commonly adopted statistical indicators: the minimum chi-square ($\chi^2_{\text{min}}$), the Akaike Information Criterion (AIC), and the Bayesian Information Criterion (BIC). These criteria allow us to compare the goodness of fit of each model while penalizing model complexity.

The AIC is defined as
\begin{equation}
    \text{AIC} = \chi^2_{\text{min}} + 2k,
\end{equation}
where $k$ is the number of free parameters in the model while the BIC is given by
\begin{equation}
    \text{BIC} = \chi^2_{\text{min}} + k \ln N,
\end{equation}
where $N$ is the number of data points used in the analysis. For both AIC and BIC, lower values indicate a more favorable model.

To compare models with $\Lambda$CDM, we define the relative differences as
\begin{align}
    \Delta \chi^2 &= \chi^2_{{\text{min}}({\Lambda \text{CDM}})} - \chi^2_{{\text{min}}({\text{model}})}, \\
    \Delta \text{AIC} &= \text{AIC}_{\Lambda \text{CDM}} - \text{AIC}_{\text{model}}, \\
    \Delta \text{BIC} &= \text{BIC}_{\Lambda \text{CDM}} - \text{BIC}_{\text{model}}.
\end{align}    
     
The interpretation of $\Delta$AIC or $\Delta$BIC is as follows: if $0 \leq \Delta X \leq 2$, the evidence is weak and both models are statistically equivalent; if $2 < \Delta X \leq 6$, the evidence is positive in favor of the model with the higher $\Delta$; if $6 < \Delta X \leq 10$, the evidence is strong; and if $\Delta X > 10$, the evidence is very strong. Here, $X$ represents either AIC or BIC.

The results of the model comparison are presented in Table~\ref{tab_model_comparison}. The $\Lambda$CDM model serves as the baseline, and all differences are computed with respect to it.


\begin{table*}[ht]
    \centering
    \renewcommand{\arraystretch}{1.5}
    \begin{tabular}{|l|c|c|c|c|c|c|}
        \hline
        \textbf{Model} & $\chi^2_{\text{min}}$ & $\Delta \chi^2$ & AIC & $\Delta$AIC & BIC & $\Delta$BIC \\
        \hline
        $\Lambda$CDM & 426.8667 & 0.0000 & 430.8667 & 0.0000 & 437.1999 & 0.0000 \\ \hline
        Hu-Sawicki   & 415.9069 & 10.9598 & 421.9069 & 8.9598 & 432.9059 & 4.2940 \\ \hline
        Starobinsky  & 409.3787 & 17.4880 & 415.3787 & 15.4880 & 426.3777 & 10.8222 \\ \hline
        Arctanh      & 415.9072 & 10.9595 & 421.9072 & 8.9595 & 432.9062 & 4.2937 \\
        \hline
    \end{tabular}
    \caption{Comparison of $\Lambda$CDM with modified gravity models using $\chi^2_{\text{min}}$, AIC, and BIC for joint observational datasets (CC + Union 3.0 + DESI-2025 + GRBs + CMB). The differences $\Delta \chi^2$, $\Delta$AIC, and $\Delta$BIC are defined relative to the $\Lambda$CDM model.}
    \label{tab_model_comparison}
\end{table*}

From the table, we observe that the Starobinsky model yields the lowest $\chi^2_{\text{min}}$, AIC, and BIC values, with $\Delta$AIC = 15.4880 and $\Delta$BIC = 10.8222, indicating very strong evidence in its favor relative to $\Lambda$CDM. The Hu-Sawicki and Arctanh models also outperform $\Lambda$CDM, with $\Delta$AIC values of 8.9598 and 8.9595, respectively, providing strong support according to the AIC. However, their $\Delta$BIC values of approximately 4.29 indicate only positive evidence over $\Lambda$CDM. Therefore, among the models considered, the Starobinsky model is the most statistically favored, followed by Hu-Sawicki and Arctanh.}


\section{Discussion and Conclusion}

In this work, we test the viability of $f(R)$ gravity as an alternative to dark energy by considering three theoretically motivated models: Hu-Sawicki, Starobinsky, and ArcTanh. For this, we combine the low-redshift probes (Cosmic Chronometers and Union~3.0 SNe) with high-redshift GRBs, CMB and the latest DESI~2025 BAO measurements.  We constrain the model parameters through Markov Chain Monte Carlo (MCMC) analysis. Our multi-probe approach provides distinct observational signatures for each modified gravity scenario while providing robust comparisons to standard cosmology. The results demonstrate how current cosmological datasets can discriminate between competing theories of late-time acceleration. \\

Our main conclusions are listed below:
\begin{itemize}
    \item \textbf{Hu-Sawicki Model:} In this model, the best fit value of Hubble constant obtained from each observations as well from joint analysis are showing good agreement with late-time probes but mild tension with early-universe measurements. For instance, the Hubble constant $H_0 = 68.921^{+0.186}_{-0.209}~\mathrm{km~s^{-1}~Mpc^{-1}}$ falls between the values inferred from early and late-universe observations. Cosmic Chronometers provide a lower estimate of $66.074^{+5.057}_{-4.357}~\mathrm{km~s^{-1}~Mpc^{-1}}$, while supernovae favor a higher value of $74.823^{+3.777}_{-3.159}~\mathrm{km~s^{-1}~Mpc^{-1}}$. Our joint analysis indicates two key results. Firstly, a statistically significant preference for modified gravity beyond $\Lambda$CDM is observed at the $99\%$ confidence level, which emerges only when combining all datasets. Secondly, the matter density constraints are found to be remarkably consistent across all observational probes. The combined analysis provides tighter constraints than any individual dataset, demonstrating how multi-probe synergies can reveal subtle gravitational effects while maintaining parameter consistency. These results suggest $f(R)$ modifications may simultaneously explain cosmological tensions and preserve standard matter content.

    \item \textbf{Starobinsky Model:} This model shows the constraints on  Hubble constant ($H_0$) closer to Planck values and a matter density ($\Omega_{m0}$) slightly higher than standard $\Lambda$CDM. The best fit value of $b$ indicates that the Cosmic Chronometers data alone shows no significant deviation from $\Lambda$CDM, while other observations (DESI~2025 BAO and supernovae) strongly favor modified gravity. Current datasets provide relatively weak constraints on the parameter $b$, with joint analysis showing the strongest preference for deviations. This suggests that while the Starobinsky model may help resolve the well-known cosmological tensions including Hubble tension, present observations lack the precision to conclusively determine its gravity modification strength across all redshift ranges.
  
    \item \textbf{ArcTanh Model:} This model produces cosmological parameters strikingly similar to $\Lambda$CDM, with a Hubble constant ($H_0$) consistent with local measurements and a matter density ($\Omega_{m0}$) showing exceptional consistency across different observational probes. While individual data alone remains fully compatible with standard cosmology, the combination with other datasets reveals a 99\% confidence level preference for modified gravity. The current constraints on the modification parameter $b$ demonstrate how multi-probe analysis can uncover subtle gravitational effects that remain hidden in individual observations, though the similar behavior to Hu-Sawicki model suggests potential parameter degeneracies that future, more precise measurements may help resolve.

    \item Further, we estimate the value of the transition redshift $(z_t)$ for each modified gravity model using different datasets. The transition redshift indicates the epoch at which the universe experienced a shift from decelerated to accelerated expansion, which offers an important test for cosmological models. In the standard $\Lambda$CDM model, $z_t$ typically falls within the range $0.6 - 0.8$, depending on the specific values of cosmological parameters. The values obtained for the Hu-Sawicki, Starobinsky, and ArcTanh models across different datasets show a general consistency with this range, though with some variations. Please refer to Table \ref{tab_transition_redshift} and Fig. \ref{fig_transition_redshift}. The Hu-Sawicki model provides $z_t \approx 0.55 - 0.66$, aligning closely with the lower bound of $\Lambda$CDM predictions. The Starobinsky model, with $z_t$ values reaching as high as $0.763$ (DESI~2025), suggests a slightly delayed transition compared to $\Lambda$CDM. Meanwhile, the ArcTanh model offers $z_t$ values similar to Hu-Sawicki, but with slightly lower bounds, indicating minimal deviation from the standard paradigm. When joint analyses are performed with datasets (CC + Union 3.0 + DESI + GRBs + CMB), the constraints become tighter, and the results converge around $z_t \approx 0.62 - 0.77$. These results suggest that while modified gravity models remain viable, their transition redshifts are largely compatible with $\Lambda$CDM.

\begin{table*}[!ht]
    \setlength{\tabcolsep}{10pt} 
    \centering
\renewcommand{\arraystretch}{1.5}

\begin{tabular}{|l|c|c|c|}
        \hline
        \multirow{2}{*}{Dataset} & \multicolumn{3}{c|}{Transition Redshift ($z_t$)} \\
        \cline{2-4}
        & Hu-Sawicki Model & Starobinsky Model & ArcTanh Model \\
        \hline
        CC & $0.554^{+0.173}_{-0.269}$ & $0.578^{+0.143}_{-0.134}$ & $0.547^{+0.160}_{-0.278}$ \\\hline
        Union 3.0 & $0.623^{+0.159}_{-0.192}$ & $0.676^{+0.113}_{-0.120}$ & $0.636^{+0.193}_{-0.204}$ \\\hline
        DESI~2025 & $0.675^{+0.031}_{-0.035}$ & $0.762^{+0.072}_{-0.104}$ & $0.659^{+0.022}_{-0.028}$ \\\hline
        CC + Union 3.0 + DESI-2025 + GRBs + CMB & $0.638^{+0.012}_{-0.013}$ & $0.763^{+0.059}_{-0.085}$ & $0.639^{+0.009}_{-0.011}$ \\
        \hline 
    \end{tabular}
    \caption{Best-fit values of transition redshift parameter at the 68\% confidence level for the Hu-Sawicki, Starobinsky, and ArcTanh models from individual and joint analyses of observational datasets.}
    \label{tab_transition_redshift}
\end{table*}
\begin{figure}[!htbp]  
    \centering
    \includegraphics[width=0.5\textwidth]{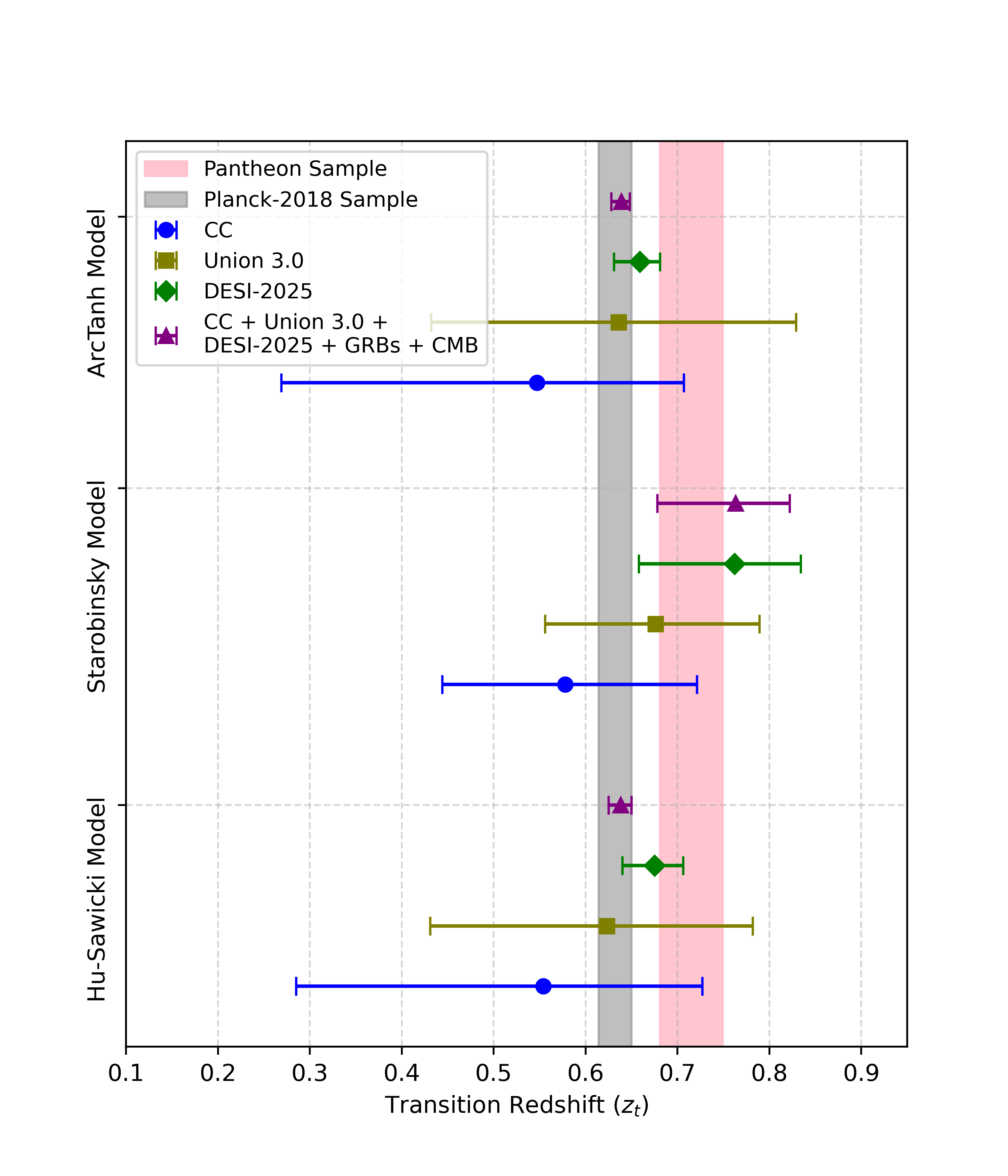}  
    \caption{Transition redshift values for different datasets across modified gravity models. The plot shows error bars for various datasets, including CC, Union 3.0, DESI~2025, and the combined dataset (CC + Union 3.0 + DESI + GRBs + CMB). Additionally, bands for the Pantheon Sample (with only statistical errors) and Planck-2018 ($\Lambda$CDM) are highlighted.}
     \label{fig_transition_redshift}
\end{figure}    
                   
\end{itemize}         
    
{Further, to evaluate model performance beyond parameter estimation, we perform a comparative analysis of the models using standard statistical tools: the minimum chi-square ($\chi^2_{\text{min}}$), Akaike Information Criterion (AIC), and Bayesian Information Criterion (BIC). This analysis allows us to quantitatively assess the relative support that the data provide for each cosmological model. The results demonstrate that all three modified gravity models offer a better fit to the data than the standard $\Lambda$CDM model. Among them, the Starobinsky model shows the strongest statistical support, with the lowest $\chi^2_{\text{min}}$, AIC, and BIC values. Specifically, it yields $\Delta \text{AIC} = 15.49$ and $\Delta \text{BIC} = 10.82$, indicating very strong evidence in favor of this model over $\Lambda$CDM. The Hu-Sawicki and Arctangent models also outperform $\Lambda$CDM, each with $\Delta \text{AIC} \approx 8.96$ and $\Delta \text{BIC} \approx 4.29$, corresponding to strong and positive levels of support, respectively. These findings suggest that modified gravity theories such as $f(R)$ models can provide viable and statistically preferred alternatives to the standard cosmological model in explaining the current observational data.}
       
The $\Lambda$CDM model remains the leading framework for describing cosmic evolution, yet modified gravity theories continue to be viable alternatives. Our analysis shows that the Hu-Sawicki and ArcTanh models exhibit only minor deviations from $\Lambda$CDM, making them nearly indistinguishable within current observational limits. The Starobinsky model, however, presents a slightly different perspective on cosmic acceleration, hinting at potential extensions to general relativity. These results suggest that while modified gravity theories can mimic standard cosmology, their unique signatures might only become evident with more precise data.        

Joint analyses significantly improve constraints on model parameters by leveraging multiple independent datasets, including Hubble parameter measurements, Supernovae, BAO, GRBs, and CMB. This approach provides tighter constraints on $H_0$, $\Omega_{m0}$, and the deviation parameter $b$, thereby improving the reliability of $f(R)$ gravity models and reducing degeneracies in parameter estimation. Future observations from high-precision surveys such as Euclid, LSST, and JWST will be essential in further constraining these models. Additionally, large-scale structure formation and gravitational wave observations could provide new avenues to test deviations from general relativity, helping to clarify whether $f(R)$ gravity can serve as a complete alternative to the cosmological constant paradigm. \\ 



\section*{Data Availability}              
\noindent This research did not yield any new data.
      
\section*{Conflict of Interest} 
\noindent The authors declare no conflict of interest.

\section*{Acknowledgments} 
We are grateful to the anonymous reviewer for their pertinent and enlightening suggestions, which greatly improved the paper. We also thank Dr. Jie Zheng for her constructive discussions on statistical model comparison techniques. Darshan would like to acknowledge RTM Nagpur University and Symbiosis Institute of Technology, Nagpur, to carry out some part of this work. He is supported by the Startup Research Fund of the Henan Academy of Sciences under Grant number 241841219. PKD and SR would like to acknowledge Inter-University Centre for Astronomy and Astrophysics (IUCAA), Pune, India for providing them Visiting Associateship under which a part of this work was carried out.

Fengge Zhang is supported by the National Natural Science Foundation of China under the Grants number 12305075, the Startup Research Fund of Henan Academy of Science under Grant number 241841223 and Joint Fund for Scientific and Technological Research of Henan Province under Grant number 235200810101.    

In this work some of the figures were created with \textbf{{\texttt{GetDist}}}~\cite{2019arXiv191013970L}, \textbf{ {\texttt{numpy}}}~\cite{numpy}  and \textbf{{\texttt{matplotlib}}}~\cite{matplotlib} Python modules and to estimate parameters we used the publicly available MCMC algorithm  \textbf{ {\texttt{emcee}}} \citep{emcee}. 

 
\bibliographystyle{apsrev4-1}
\bibliography{references_drsn}

\end{document}